\documentclass[aip,graphicx]{revtex4-1}

\usepackage{graphicx}
\usepackage{amsmath}
\usepackage{amssymb}

\begin{document}


\title{Self-organisation of semi-flexible rod-like particles}



\author{Bart de Braaf}
\affiliation{Department of Applied Physics, Eindhoven University of Technology, PO Box 513, 5600\,MB Eindhoven, The Netherlands.}

\author{Mariana Oshima Menegon}
\email[Corresponding author: ]{m.oshima.menegon@tue.nl}
\affiliation{Department of Applied Physics, Eindhoven University of Technology, PO Box 513, 5600\,MB Eindhoven, The Netherlands.}

\author{Stefan Paquay}
\altaffiliation[Present address: ]{Department of Physics, Brandeis University, 415 South St., Waltham, MA, USA.}
\affiliation{Department of Applied Physics, Eindhoven University of Technology, PO Box 513, 5600\,MB Eindhoven, The Netherlands.}

\author{Paul van der Schoot}
\affiliation{Department of Applied Physics, Eindhoven University of Technology, PO Box 513, 5600\,MB Eindhoven, The Netherlands.}
\affiliation{Instituut voor Theoretische Fysica, Universiteit Utrecht, Princetonplein 5, 3584\,CC Utrecht, The Netherlands.}


\date{\today}

\begin{abstract}
We report on a comprehensive computer simulation study of the liquid-crystal phase behaviour of purely repulsive, semi-flexible rod-like particles. For the four aspect ratios we consider, the particles form five distinct phases depending on their packing fraction and bending flexibility: the isotropic, nematic, smectic A, smectic B and crystal phase. Upon increasing the particle bending flexibility, the various phase transitions shift to larger packing fractions. Increasing the aspect ratio achieves the opposite effect. We find two different ways how the layer spacing of the particles in the smectic A phase may respond to an increase in concentration. The layer spacing may either decrease or increase depending on the aspect ratio and flexibility. For the smectic B and the crystalline phases, increasing the concentration always decreases the layer spacing. We find that the layer spacing jumps to a larger value on transitioning from the smectic A to the smectic B phase.
\end{abstract}


\maketitle 

\section{Introduction}

Rod-like colloidal particles, DNA strands, carbon nanotubes, and filamentous viruses have in common that, if dispersed in a fluid at sufficiently high concentrations, they exhibit various kinds of liquid-crystalline phase.
This is because with increasing concentration, the dispersion runs out of free volume leading to increasingly ordered states.
This class of material is usually referred to as lyotropic liquid crystals, which sets them apart from thermotropic liquid crystals because the driving force is not energy but, in essence, entropy.
This was first recognised by Lars Onsager in his seminal paper describing the isotropic-to-nematic phase transition of cylindrical particles interacting via a hard-core repulsive interaction \cite{Onsager1949}.
In agreement with experiment, the theory predicts the transition to occur at a volume fraction that decreases inversely proportional to the aspect ratio of the particles.
The impact of particle bending flexibility on the isotropic-nematic transition was first investigated theoretically by Khokhlov and Semenov more than thirty years later \cite{Khokhlov1981,Khokhlov1982}, a decade after that others investigated how flexibility impacts upon the nematic-columnar and the nematic-smectic A transition \cite{Selinger1991i,Selinger1991ii,Tkachenko1996,vdSchoot1996}. 

Over the past few decades, interest in lyotropic liquid crystals has increased significantly, in part because of potential applications and in part because of the development of well-controlled model particles \cite{Vroege1992,Kuijk2012}.
Indeed, lyotropic liquid crystals have been intensively investigated experimentally \cite{Grelet2008dyn,Grelet2008hex,Grelet2014,Grelet2016}, theoretically \cite{Odijk1985,Odijk1986,Chen1993,Hidalgo2005,Shundyak2006} and with the aid of computer simulations \cite{Dijkstra1995,McGrother1996,Bladon1996,Cinacchi2008,Naderi2013,Naderi2014,Egorov2016}.
In spite of this, our understanding of the isotropic and nematic phases is most comprehensive, whilst that of the others remains much more sketchy.
In particular, how flexibility and aspect ratio impact upon the other liquid crystal transitions have received much less attention.
Here, we aim to fill in this gap from the perspective of computer simulations, in particular because these are much more difficult to address theoretically. One reason is that the second virial approximation, which allowed Onsager to accurately describe the isotropic-nematic transition, no longer holds at densities where the smectic and columnar phases appear. Another reason is translation-rotation coupling, which makes density functional and integral equation theories virtually intractable \cite{vanRoij1995,vdSchoot1996,Hidalgo2005}.  

We extend earlier simulation studies on semi-flexible chains by covering a larger range in persistence length, aspect ratio, and particle numbers, and investigate more comprehensively the microscopic structure of the liquid crystalline and crystalline phases. In agreement with theory and simulation, we find that particles with longer aspect ratio support over a larger concentration range and a broader range of bending flexibilities liquid-crystalline states. This is particularly true for the nematic and the smectic A phase. We find that the stability of the smectic B and crystalline phases, recently both found experimentally in colloidal systems, to be less sensitive to both aspect ratio and flexibility, at least for the ranges investigated. The aspect ratio of our particles varied between 6 and 11 while the ratio of the bare contour length and persistence length varied between 0.05 and 0.5. For these aspect ratios and persistence lengths all phase transitions are either second or weakly first order, except the transition between smectic A and smectic B that clearly is first order. The difference in behaviour of the smectic A and B phases expresses itself most clearly in how the layer spacing responds to increases in density. For the smectic B phase, the layer spacing always decreases with increasing density. This is not so for the smectic A phase, where depending on aspect ratio and persistence length it may in- or decrease, depending on whether the increase of the particles density is translated into reduced layer spacing and/or increased in-layer density.  

The remainder of this paper is structured as follows. In Section \ref{sec:met} we describe our model particles that we construct from overlapping, mutually repulsive bead-spring chains. We also make explicit our simulation protocol and explain how we identify the various liquid crystal phases in our simulation data. In Section \ref{sec:res} we present the phase diagrams we obtain and show how the aspect ratio and flexibility influence the phase transitions. In Section \ref{sec:length} we describe the influence of the phase transitions on the individual particle structure and smectic layer thickness. We furthermore present a simple model based on Onsager theory that explains the changes in the particle length as a function of density in the isotropic and nematic phases in Appendix \ref{appendx}. Finally, in Section \ref{sec:con}, we present our most important conclusions.

\section{Methods and analysis }
\label{sec:met}

\begin{figure}
 \includegraphics[scale=1]{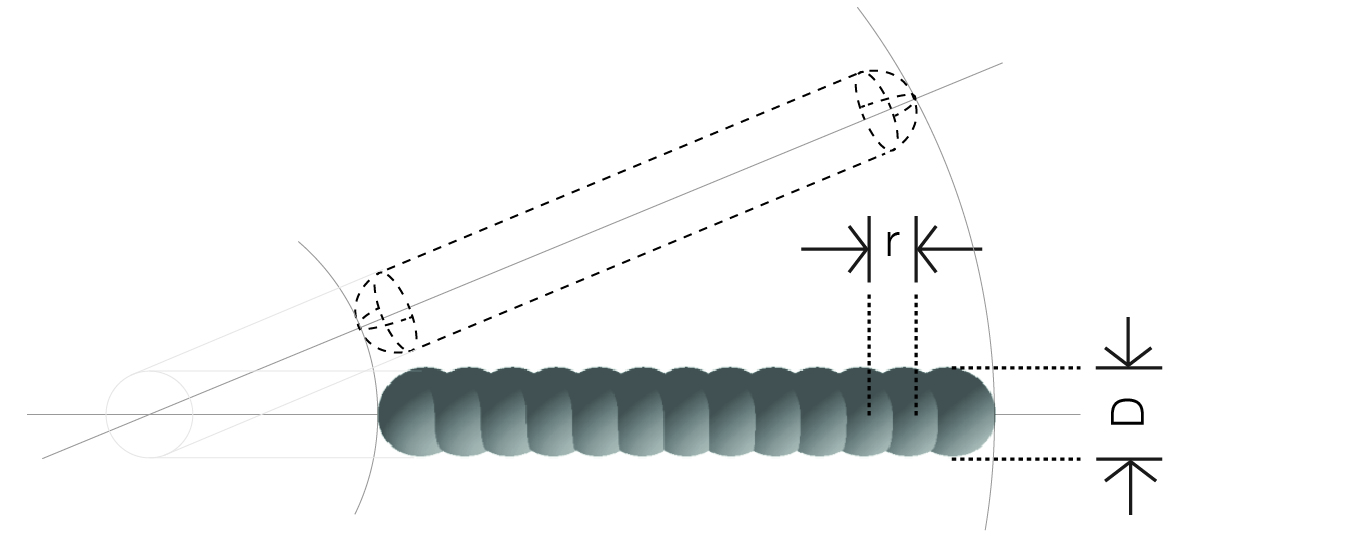}%
 \caption{\label{chain}Schematic representation of the semi-flexible rod-like particles. They are modelled as a bead-spring chain with diameter $D$. The beads partially overlap for a smoother surface, allowing closer comparison with spherocylinders (dashed schematic). The rest distance $r$ of the harmonic bond potential corresponds to half the bead diameter.}%
 \end{figure}

To study the equilibrium properties of semi-flexible rod-like particles, we perform MD simulations on 4608 bead-spring chains consisting of $n$ beads of mass $m$ using the software package LAMMPS \cite{Plimpton1995}. Within a chain, consecutive pairs of bead interact via a harmonic potential with force constant $\kappa$. The rest distance $r$ between their centres of mass corresponds to half the bead diameter $D$. See Figure \ref{chain}. Hence, the beads partially overlap in order to provide a smoother particle surface and to prevent biased stacking between the rod-like particles in highly congested phases. A harmonic bending potential with bending constant $\kappa_\theta$ is assigned to consecutive bonds between beads to model bending stiffness. Except for the nearest neighbour beads along the same chain, all beads interact via a truncated and shifted Lennard-Jones (LJ) potential with effective diameter $\sigma=D$ and interaction strength $\epsilon=k_BT$, where $k_B$ is the Boltzmann constant and $T$ the absolute temperature. We truncate the potential at the centre-to-centre distance of $2^{1/6}\sigma$ to make the potential purely repulsive. 
 
We perform isobaric-isothermal (NPT) simulations at various pressures. To control temperature and pressure in our simulations, we employ the Nos\'{e}-Hoover thermostat and barostat. The thermal energy $k_BT$ is our reference energy unit. The barostat can adjust the rectangular simulation box dimensions independently, which allows relaxation to the correct layers spacings in the smectic and crystalline phases. Each simulation runs for 20000 time units. Our time unit is set by $(m/k_BT)^{1/2}\sigma$. The relaxation times for temperature and pressure are $0.01$ and $0.1$ time units corresponding to about 10 and 100 time steps, respectively. Approximately 200 configurations of every run are stored, \emph{i.e.}, one every 100 time units.
 
The initial configuration is that of the crystal phase, with all rod-like particles perfectly aligned, AAA stacked in 16 layers with $18\times16$ particles each, \emph{i.e.}, the layers are identical copies shifted along the director, and in each layer there is perfect hexagonal ordering. That the initial box is very elongated is sensible because the particles themselves have a large aspect ratio. In the isotropic phase the box elongation relaxes and on average becomes isometric albeit that the box shape fluctuates considerably, in particular near the isotropic-to-nematic phase transition. In the nematic phase the box can become very much more elongated than the initial elongation. In the smectic and crystalline phase the box anisometry remains roughly equal to the initial one. If in our simulations one box dimension drops below about one particle length we discard the run.

We set the elastic constant $\kappa$  at a large value of $100\,k_B T/\sigma^2$ to ensure minimal entropic stretching of the bonds. In other words, the average bond length is very close to the rest bond length of one-half $\sigma$. In our simulations we allow for chains consisting of $n=13,\;15,\;17,\; \text{and}\; 21$ beads per chain. The corresponding aspect ratios $L_0/D$ in the limit of zero pressure we find to be $L_0/D=6.46,\;7.54,\;8.62,\;\text{and}\;10.77$. For every aspect ratio we vary the bending stiffness $\kappa_{\theta}$ to obtain a series of ratios of the contour length $L_0$ and the persistence length $L_p$, $L_0/L_p=0.05,\;0.1,\;0.2,\;0.3,\;0.4\;\text{and}\;0.5$. The persistence length we calculated from the equality $L_p=\kappa_{\theta}r/k_BT$, valid for an infinitely large number of beads and $\kappa_{\theta}r^2/k_BT\gg 1$ \cite{Naderi2014}.

To calculate the corresponding volume fraction $\phi$ of any given configuration, we take the equilibrated volume of the simulation box $V$ for a given pressure $P$, and define the volume occupied by the particles as the fixed number of chains $N$ in the system times the occupied volume $v_0$ by each chain, so $\phi=Nv_0/V$. The volume of a chain we approximate by taking a spherocylinder with volume $v_0=\pi{D^3}/6+\pi{D^2}L_0/4$.

\begin{figure}
 \includegraphics[scale=.4]{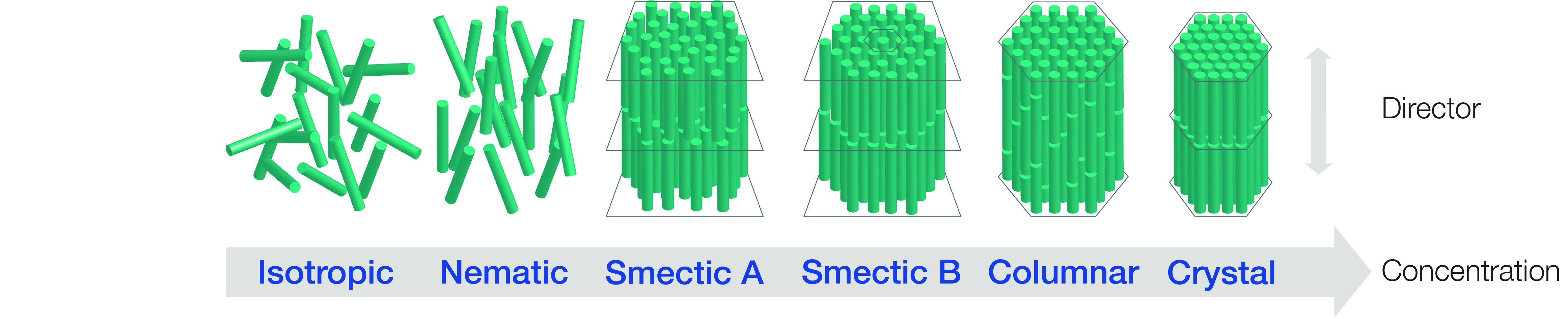}%
 \caption{\label{phaseScheme}Schematic representation of the liquid crystal phases found in aqueous dispersions of the rod-like fd virus \cite{Grelet2014}. The phase sequence with increasing concentration is isotropic, nematic, smectic A, smectic B, columnar, and crystal. The double pointed arrow indicates the preferential direction of the aligned particles, the director.}%
 \end{figure}

The equilibrium configurations stored are used to calculate the usual order parameters and the pair correlation functions of the collection of particles. The order parameters quantify (1) the degree of orientation of the chains, given by the nematic order parameter $S_2$; (2) the organisation in layers perpendicular to the director, given by the smectic order parameter $\tau$; and (3) the hexagonal ordering of the closest neighbours within the same layer, which is described by the bond order parameter $\psi_6$ \cite{deGennes,Polson1997,Nelson1979}. With these order parameters, the isotropic, nematic, smectic A, columnar, and smectic-B or crystal phases can be identified and distinguished. All these phases are schematically represented in Figure \ref{phaseScheme}.
 
\begin{figure}
 \includegraphics[scale=.3]{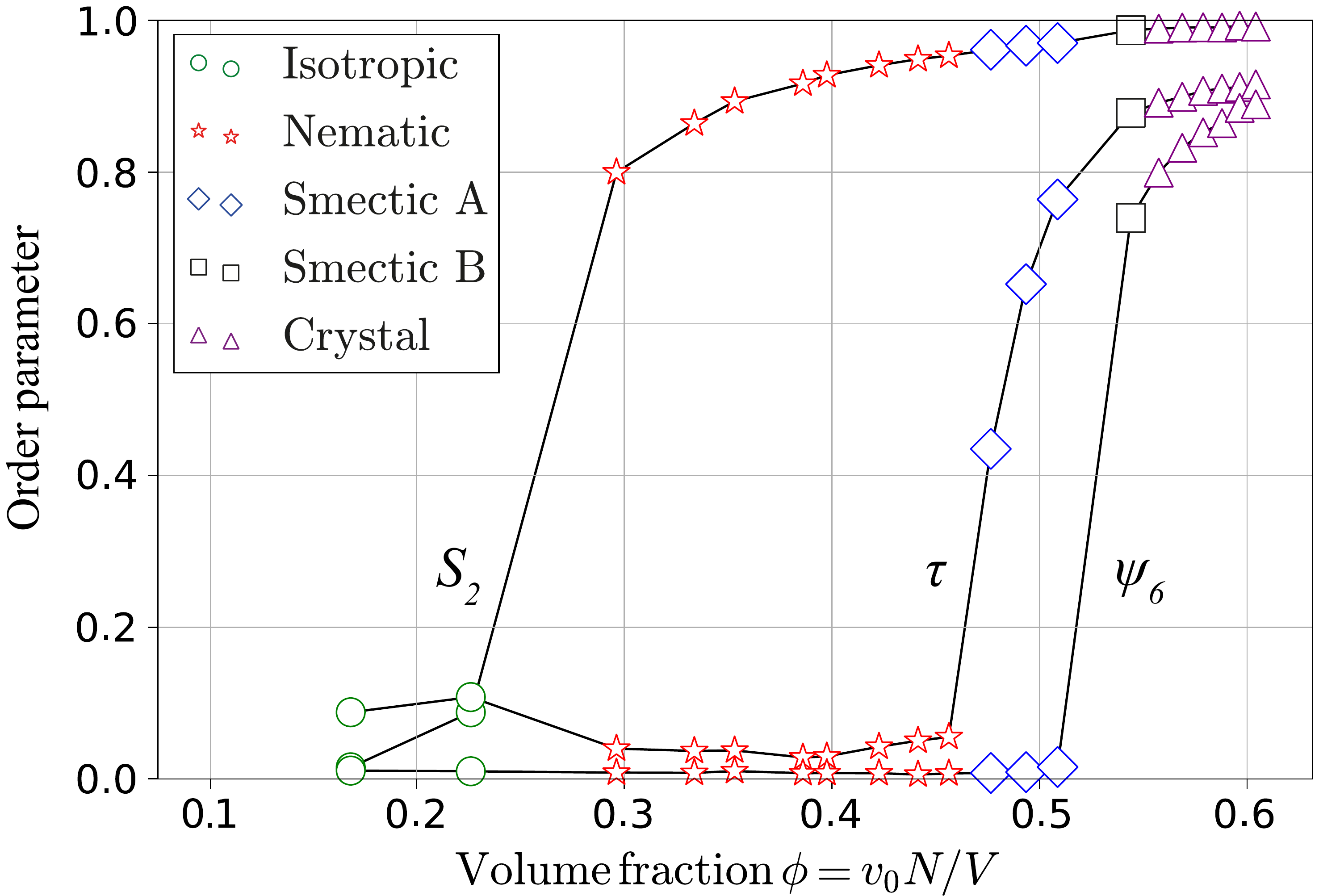}%
 \caption{\label{orderOP}Order parameters as a function of the volume fraction $\phi$ for aspect ratio $L_0/D=10.77$, flexibility $L_0/L_p=0.05$. Respectively the nematic order parameter $S_2$, the smectic order parameter $\tau$, and the bond order parameter $\psi_6$.}%
 \end{figure}

\begin{figure}
 \includegraphics[scale=.25]{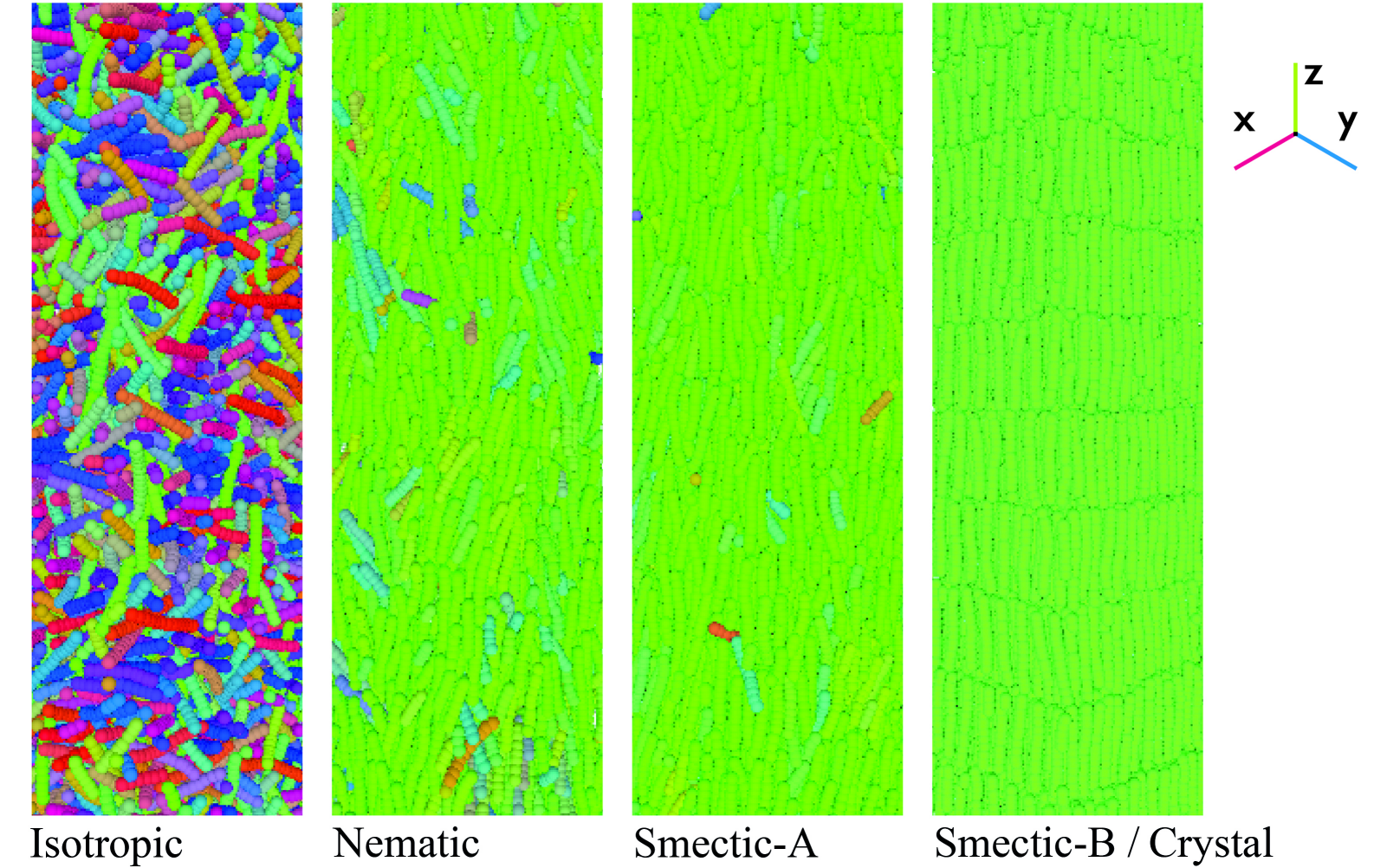}%
 \caption{\label{phases}Snapshots representing the arrangement of particles along the director for the different phases observed in our simulations for aspect ratio $L_0/D=6.46$ and ratio of contour length and persistence length of $L/L_p=0.1$. From left to right with increasing density: isotropic, nematic, smectic A, smectic B, and crystal.}%
 \end{figure}

\begin{figure}
 \includegraphics[scale=.38]{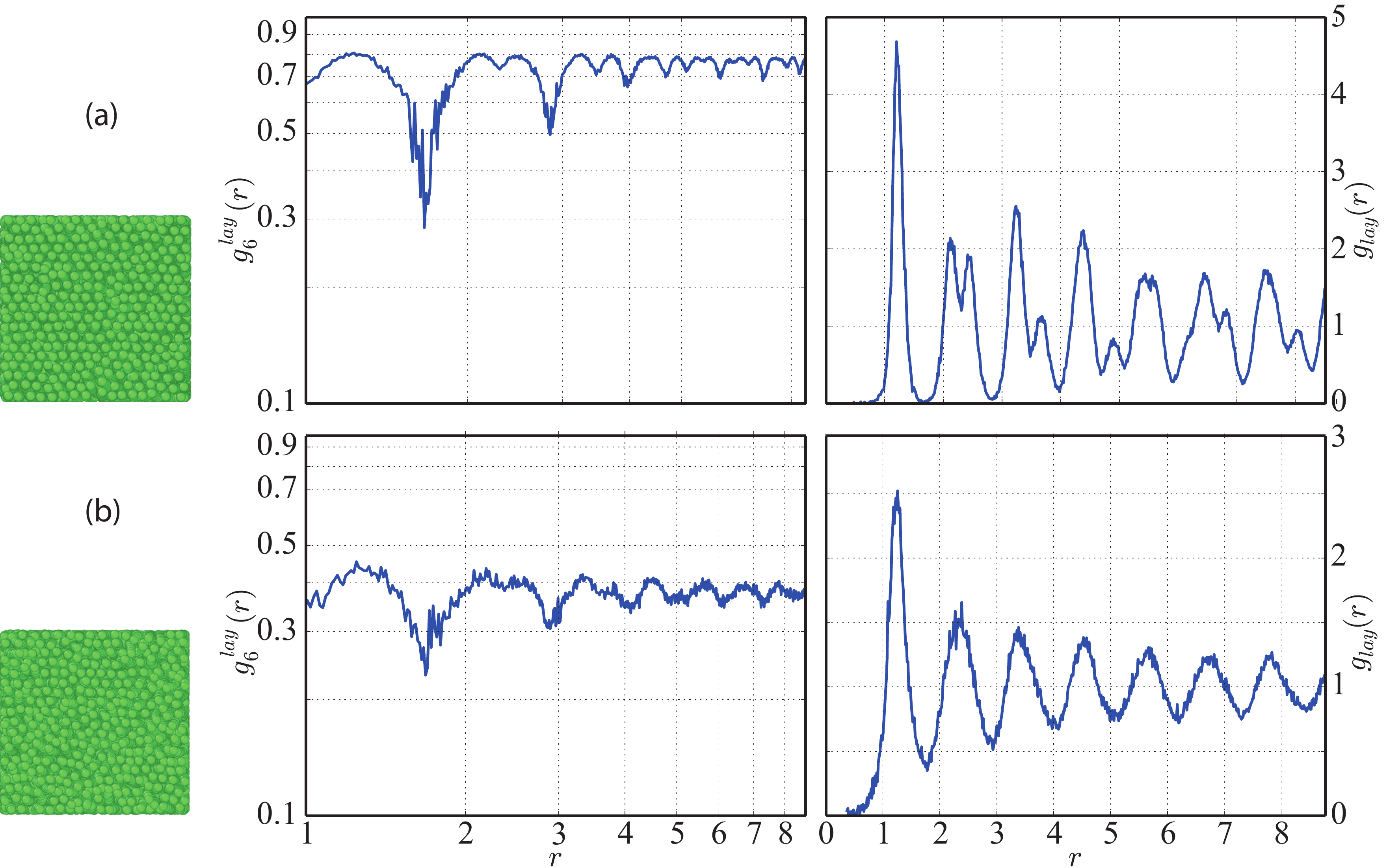}%
 \caption{\label{paarcorr}(a) The $\psi_6$ correlation function $g_6^\mathrm{lay}(r)$ (left) and the pair correlation function $g_\mathrm{lay}(r)$ (right) for particles with aspect ratio $L_0/D=10.77$, flexibility $L_0/L_p=0.5$ and volume fraction $\phi=0.63$. The pair correlation function exhibits peaks characteristic for a crystal phase and hence we identify it as such. (b) The same for a volume fraction of $\phi=0.57$. The pair correlation function does not show the characteristic crystal peaks. Hence at a volume fraction of $\phi=0.57$ the particles must in the smectic B phase. See also the main text.}%
 \end{figure}

In the isotropic phase there is only short-range correlation between the positions and between the orientations of the chains, and $S_2$ should be zero, but need not be in account of finite size effects. With the alignment of the particles in the nematic phase, the order parameter $S_2$ increases abruptly when crossing the phase boundary, so it can be readily identified. The same is true for the smectic and bond order parameters $\tau$ and $\psi_6$, allowing us to identify the smectic A phase and the smectic B or crystal phase. See Figure \ref{orderOP}. Snapshots of the various phases are given in Figure \ref{phases}. We cannot distinguish between the smectic B and crystal phase based only on calculating the correlation function $g_6^\mathrm{lay}$ based on  the order parameter $\psi_6$ due to the finite size of the system. See Figure \ref{paarcorr}. However, we can distinguish between them by considering the in-layer pair correlation function $g_\mathrm{lay}$ of the centres of mass of the chains \cite{Qi2008,Naderi2014}. 

%

As can be seen in Figure \ref{paarcorr} (a) and (b), there is a clear difference in $g_{\mathrm{lay}}(r)$ between two states at different pressures with equal magnitude of the order parameters $S_2$, $\tau$, and $\psi_6$. Figure \ref{paarcorr} (a) exhibits a split second peak in $g_{\mathrm{lay}}(r)$, a characteristic of a crystalline phase that the system with the pressure shown in (b) does not have. We therefore associate the absence of peak splitting with the smectic B phase and assume the phase transition takes place when the second peak in the pair correlation function splits. So, we use the splitting of the second peak in $g_\mathrm{lay}(r)$ as a proxy for distinguishing between the smectic B and the crystal phase. Note that the smectic B phase that we identify in (b) has a much noisier $g_6^\mathrm{lay}$ than that of the crystal phase of (a). The difference in structure of the crystal and smectic B phase is also evident from the snapshots also presented in Figure \ref{paarcorr}. 
 
In order to determine the crystal symmetry, we compare the pair correlation function $g_\mathrm{lay}$ of the centers of mass of the particles. We distinguish four cases. In the first, we calculate the pair correlation function for particles in the same layer. In the second, we consider pairs of particle in consecutive layers. In the third and fourth, the pair correlation function considers pairs of particle separated by one and two layers. We expect that all pair correlation functions must be similar for the AAA crystal structure, whilst the first and fourth should be similar for the ABC structure. We observe neither of these patterns, implying that we cannot pinpoint the exact crystal structure. A possible explanation for this is that the ordering between layers is not so well defined for semi-flexible particles. The fact that we start off from an initial AAA structure that does not seem to survive we conclude that our simulations are not kinetically trapped. 

\section{Results and discussion}
\subsection{Phase diagrams}
\label{sec:res}

\begin{figure}
 \includegraphics[scale=.25]{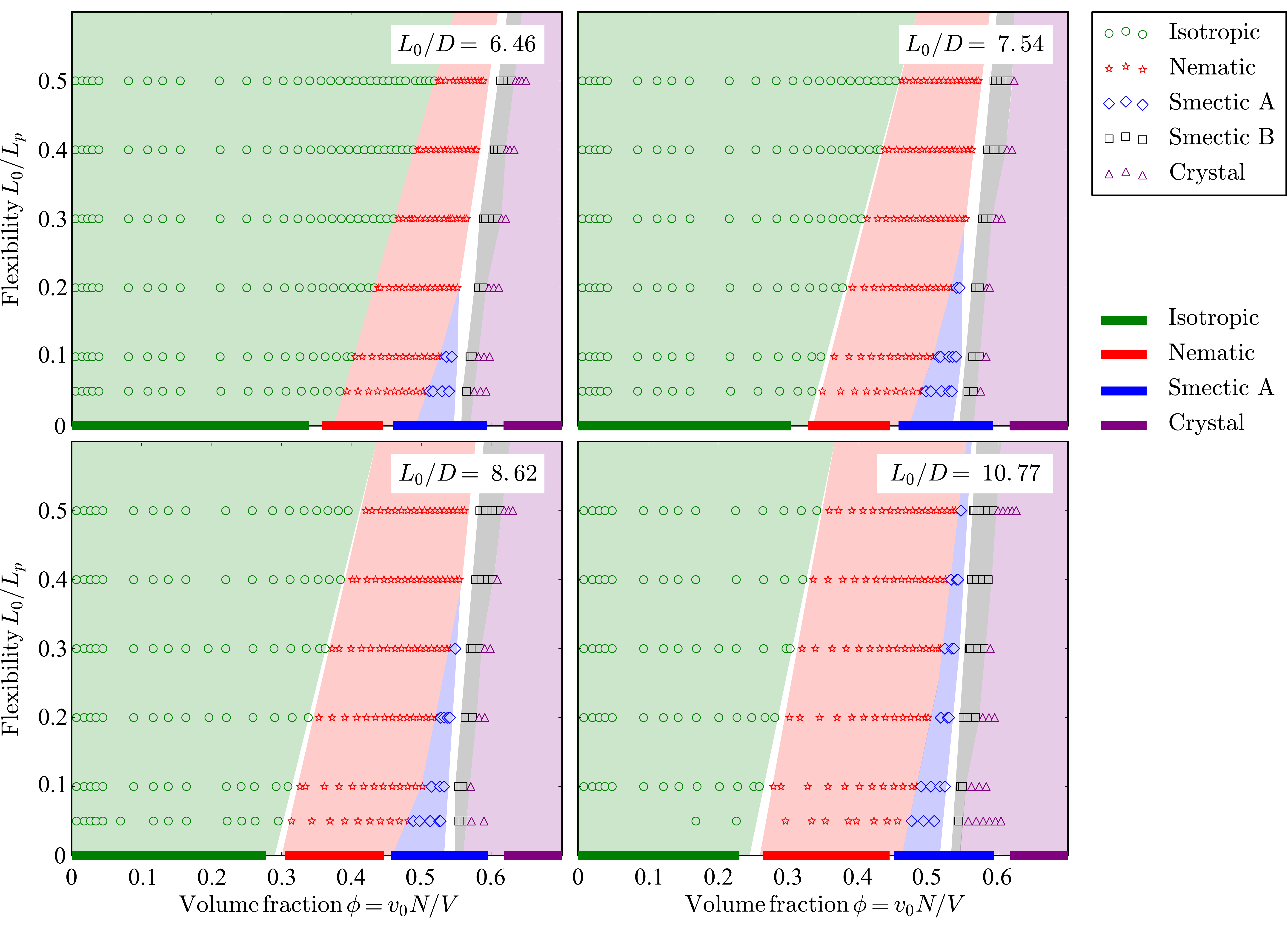}%
 \caption{\label{phasediagramLLP}Phase diagrams as a function of the volume fraction $\phi$ and flexibility $L_0/L_p$ for rods with aspect ratios $L_0/D=6.46$, 7.54, 8.62, and 10.77. Indicated are the isotropic phase (green circles), the nematic phase (red stars), the smectic A phase (blue diamonds), the smectic B  phase (black squares), and the crystal phase (purple triangles). Corresponding background colours are added to aid identifying the various phases. The isotropic-nematic and the nematic-smectic A phase transitions shift to higher volume fractions with increasing degree of flexibility. Furthermore the smectic A phase disappears above a critical, aspect-ratio dependent degree of flexibility. The smectic A-smectic B or nematic-smectic B and smectic B-crystal phase transitions shift to larger volume fractions with increasing degree of flexibility albeit that the effect is relatively weak. The bars placed at zero flexibility indicate the simulation results of Bolhuis and Frenkel for infinitely rigid, hard spherocylinders for comparison \cite{Bolhuis1997}.}%
 \end{figure}

The phase diagrams of our particles are presented in Figure \ref{phasediagramLLP}. Recall that our particles are semi-flexible, rod-like chains interacting via a soft-core, repulsive potential. We present phase diagrams as a function of volume fraction and bending flexibility, ranging from $L_0/L_p=0.05$ to 0.5 covering particles from near the rigid-rod limit to semi-flexible chains, for four aspect ratios, $L_0/D=6.46$, 7.54, 8.62, and 10.77. We distinguish between the following phases: isotropic, nematic, smectic A, smectic B and crystal (Figure \ref{phases}). For our set of parameters, we did not encounter any evidence for a columnar phase. Based on what we know on the phase behaviour of the fd virus, which does support a columnar phase, we must conclude that our particles do not have large enough aspect ratio for this phase to appear in the phase diagram \cite{Grelet2016}.

Focusing on the aspect ratio $L_0/D=6.46$ first, representing the trends observed for the other aspect ratios, Figure \ref{phasediagramLLP} tells us that all phase transitions shift to larger volume fractions with increasing flexibility. The isotropic-nematic transition increases approximately linearly with increasing degree of flexibility, which for large persistence lengths is consistent with theory and Monte Carlo \cite{Dijkstra1995}. Both the isotropic-nematic and the nematic-smectic A transition are significantly impacted upon by any bending flexibility. 
Theoretically, this has been predicted to be the case albeit that these theories are typically valid in the long-chain and/or large persistence length limits relative to the width of the particles \cite{Khokhlov1981,Khokhlov1982,Odijk1985,Odijk1986,Chen1993,Tkachenko1996,vdSchoot1996}. The result also also agrees with previous simulation by Bladon and Frenkel \cite{Bladon1996}.
We find that the smectic A phase is strongly destabilised by decreasing the chain stiffness, in line with results from earlier computer simulations by Cinacchi and Gaetani on shorter rods and for smaller box sizes \cite{Cinacchi2008}.

For values of $L_0/L_p>0.1$ we find a direct transition from the nematic to the smectic B phase, \emph{i.e.}, the smectic A phase disappears for large enough flexibilities. We notice that the transitions between the nematic and the smectic B, the smectic A and the smectic B, and the smectic B and the crystal phase are much less sensitive to changes in particle flexibility, and in fact to variations in the aspect ratio. The smectic A phase is more stable for larger aspect ratios and present in the phase diagram all flexibilities probed for the aspect ratio $L_0/D=10.77$. The transition from the smectic A or nematic to the smectic B, and that from the smectic B to the crystal phase, are only very weakly dependent on the aspect ratio and bending flexibility of the particles. This is not entirely unexpected, on the one hand because the particles in these dense phases are almost perfectly aligned, and on the other hand because the Odijk deflection length $\lambda_\mathrm{Odijk}=L_p\langle\theta^2\rangle$ turns out to be of the order the width of the particles in those phases. This implies that bending modes with smaller wavelengths cannot be suppressed and that in this limit bending flexibility should be unimportant \cite{Odijk1985}. 
Practically, this is true if the degree of alignment of particles, given by the nematic order parameter $S$, is larger than $1-(3D)/(2L_p)$. This happens to be the case for the smectic B and crystal phases for the range of flexibilities that we cover.

Our simulation results are consistent with those of Bolhuis and Frenkel for rigid, hard spherocylinders \cite{Bolhuis1997}, represented in Figure \ref{phasediagramLLP} by the bars placed at zero flexibility ($L_0/L_p\rightarrow0$). The agreement is even quantitative for less ordered phases whilst for the highly ordered phases the phase transitions in Bolhuis and Frenkel's simulations shift to larger concentrations compared to ours. There are several explanations for this. First, our rod-like chains are slightly compressible. As we shall see in the next section, excluded-volume interactions cause the chains to compress in particular in the phases where free volume become scarce, so in the denser phases.    
Second, our particles interact through a soft-core interaction while the rigid rods of Ref. \cite{Bolhuis1997} interact via a hard-core potential. Third, our simulation box is much larger than that of 1997 study of Bolhuis and Frenkel. Their particle number was at most 600 whilst in our case it is 4608, suggesting that finite size effects might also play a role in discrepancy.  

Regarding the order of the transitions, we can only confirm that the transition from the nematic or the smectic A to the smectic B phase is most definitely of first order: we observe a clear jump in the density at the pressure where the transition takes place (results not shown). 
We find the isotropic-to-nematic transition to be weakly first order, if at all, but it seems to become more strongly first order with increasing aspect ratio, to shift to lower concentrations and generally to become more stable. This is in line with the computer simulations of Bolhuis and Frenkel \cite{Bolhuis1997}. 
For the other transitions, we find that, if there are jumps, we do not have the resolution to observe them. 
The experiments of, \emph{e.g.}, Grelet \emph{et al.} on aqueous dispersions of fd virus particles, which have an aspect ratio 130, indicate that the nematic-to-smectic A transition is first order  \cite{Grelet2008dyn,Grelet2008hex,Grelet2014}. The order of the transition from smectic A to smectic B for fd virus remains unclear. Fd virus does not transition from smectic B to crystal but to a columnar phase \cite{Grelet2008hex,Grelet2014}. We hypothesise that the large aspect ratio of the viruses particles somehow stabilises the columnar phase.   

Having discussed the macroscopic (thermodynamic) properties of our particles, we next investigate in more depth how the particles and the structure of the more ordered phases respond to particle length and flexibility. Interestingly, we find that the layer spacing in the smectic A phase may increase or decrease with increasing concentration depending on flexibility and aspect ratio of the chain. This increase of the layer spacing with increasing density is counter intuitive but, as we shall see next, is somehow connected with the layer spacing.  

\subsection{Microscopic structure}
\label{sec:length}

Our first probe of the microscopic structure of the various phases is the actual contour length of the chains relative to the unperturbed contour length. This is important because our particles are not only flexible but also slightly compressible. Hence, we expect that with increasing particle density they should become shorter in order to accommodate a decreasing free volume. This can be seen as a drawback of our model particles but in fact allows us to address the question to what extent particle flexibility impacts upon the excluded volume in the isotropic phase, and, \emph{vice versa} if and how excluded volume interactions impact upon the effective particle bending flexibility. 

\begin{figure}
 \includegraphics[scale=.25]{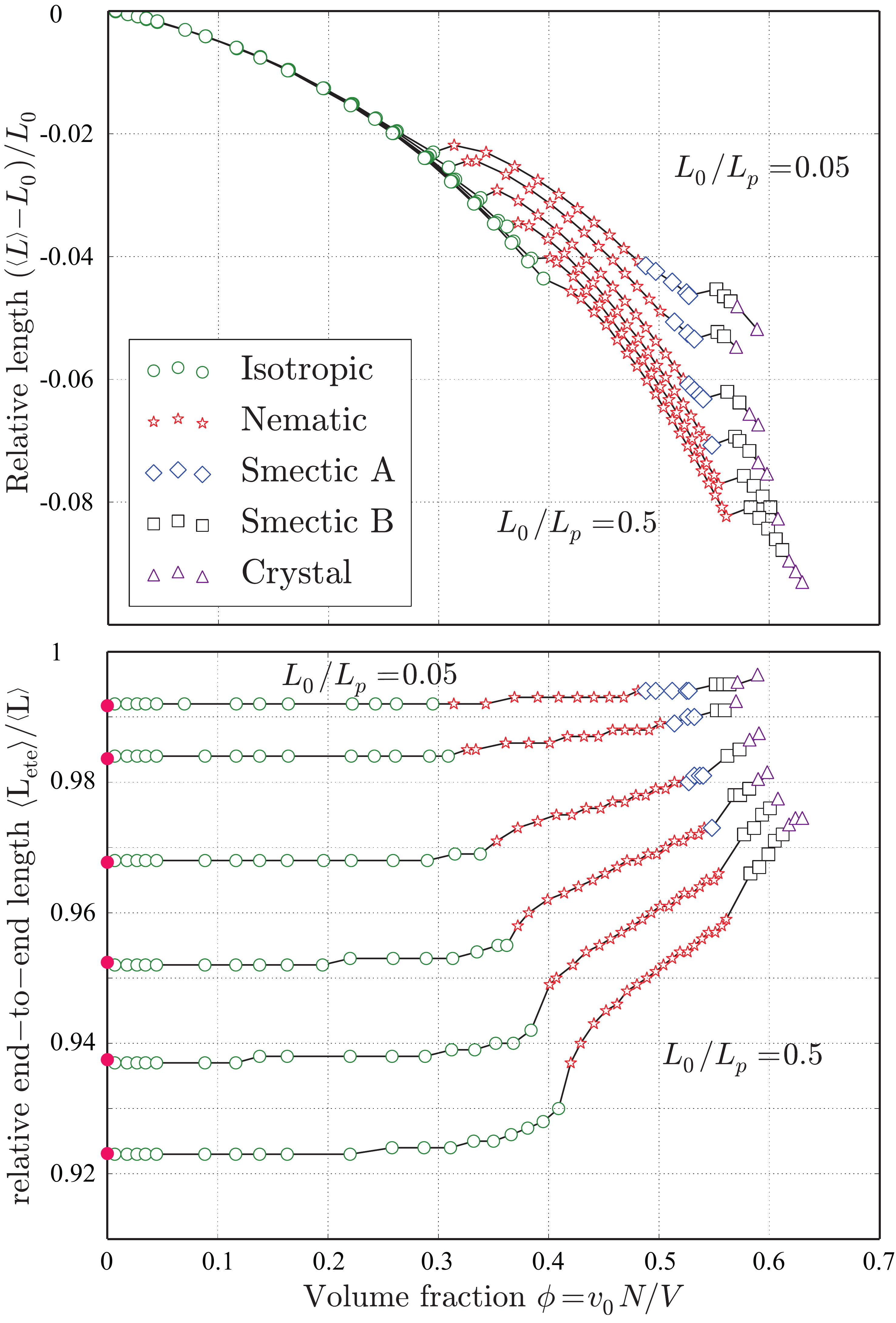}%
 \caption{\label{eteN17}The average change in contour length of the chains $(\langle{L}\rangle-L_0)/L_0$ as a function of the volume fraction $\phi$ for various aspect ratio $L_0/D = 8.62$ (top). The symbols are defined in Figure \ref{phasediagramLLP}. The compression and jumps in length are explained in the main text. The relative end-to-end lenght $\langle{L_{\mathrm{ete}}}\rangle/\langle{L}\rangle$ as a function of the volume fraction $\phi$ for aspect ratio $L_0/D = 8.62$ (bottom). Filled circles in magenta represent the prediction for the worm-like chain model.}%
 \end{figure}

In Figure \ref{eteN17}a, the contour length $\langle{L}\rangle$ is scaled to the reference contour length $L_0$ for the aspect ratio $L_0/D=8.62$ as a function of the volume fraction and the flexibility. The contour length decreases with increasing volume fraction in the isotropic phase. This decrease does not depend on particle flexibility suggesting that volume exclusion in the isotropic phase is an invariant of the particle flexibility, as has been presumed in the past \cite{Khokhlov1981,Khokhlov1982,Vroege1992,Chen1993,Kuijk2012}. We observe a small but sudden increase of the contour length at the isotropic-nematic transition, except for the most flexible chains for which the transition seems to become either second or very weakly first order. Arguably, the reason for this jump is an increased free volume caused by the alignment of the particles in the nematic phase \cite{Kuijk2012}. This confirms that the transition is first order albeit more weakly so for the more flexible chains. In the nematic phase, the contour length decreases with increasing concentration, again because of the decrease in free volume with increasing concentration. A much stronger jump we find on going from the nematic or smectic A to the smectic B phase. Simple second virial calculations presented in the Appendix \ref{appendx} confirms the observed trends for the isotropic and nematic phases, explaining also the jump in length.

In Figure \ref{eteN17}b, the end-to-end length  $\langle{L_{\mathrm{ete}}}\rangle$ is scaled to the measured contour length $\langle{L}\rangle$ for aspect ratio $L_0/D=8.62$ as a function of the volume fraction and the flexibility. In the isotropic phase, this end-to-end length apparently depends only on the ratio $L_0/L_p$. It depends weakly on the concentration except for the most flexible chains and then only near the isotropic-to-nematic phase transition. This we argue is again caused by the excluded-volume interactions not being affected by particle flexibility. Our measured values for the relative end-to-end length are in very good agreement with prediction given by the worm-like chain model, also indicated in the figure. This confirms that our estimate of the persistence length for our model chains is accurate. In the liquid crystalline phases the end-to-end distance does depend on the concentration and more so the more flexible the particles. This can straightforwardly be understood by realising that a combination of persistence and the molecular ordering field attenuates the bending fluctuations \cite{Odijk1985} . The latter becomes stronger the larger the particle density. The same is true for the remaining phase transitions as in fact we already alluded to in the previous section. 
 
\begin{figure}
 \includegraphics[scale=.25]{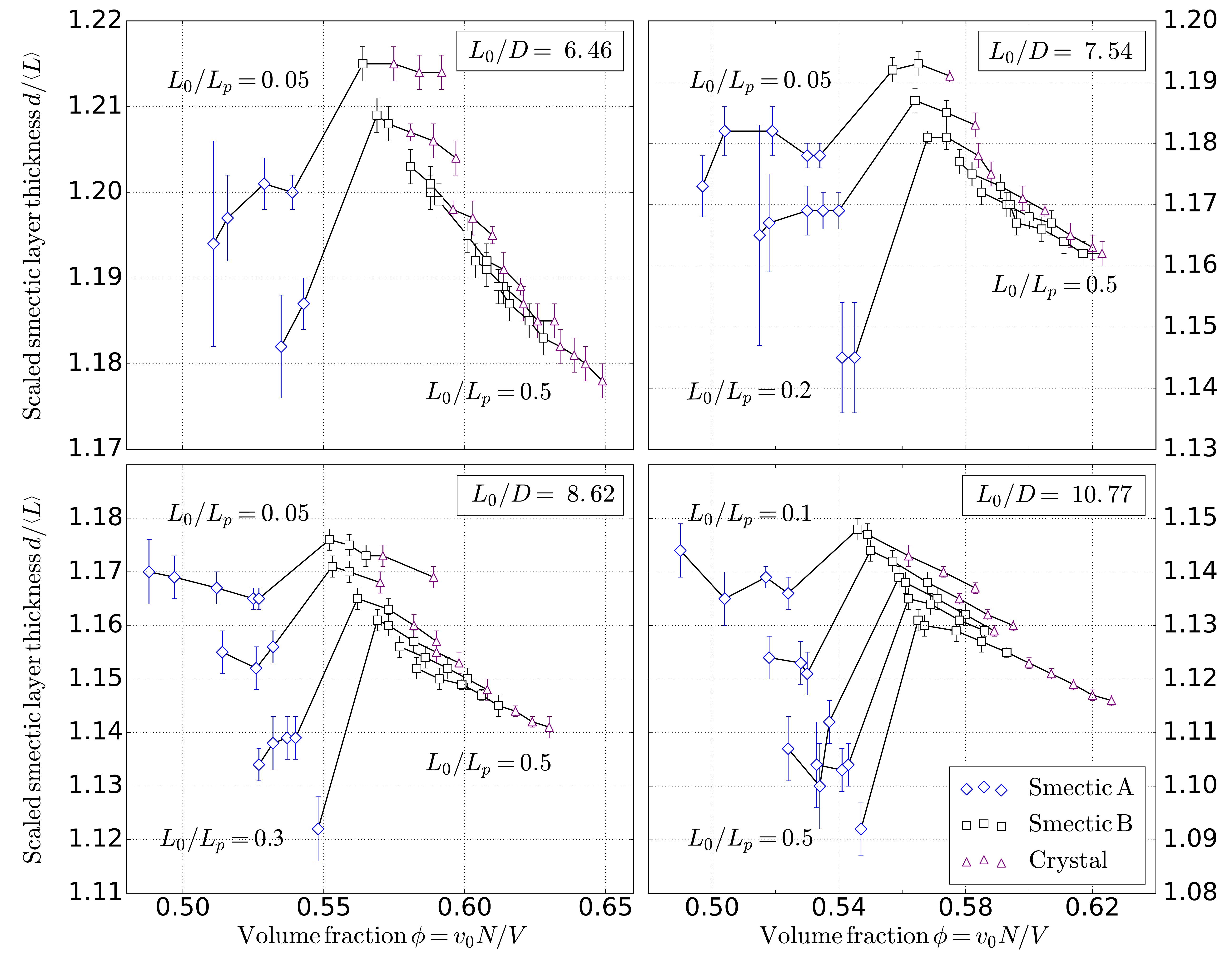}%
 \caption{\label{LayerN17}The scaled smectic layer thickness $d/\langle{L}\rangle$ versus density $\phi$ for various aspect ratios $L_0/D$. The symbols are introduced in Figure \ref{phasediagramLLP}. The smectic layer thickness is scaled to the measured average contour length of the particles at that volume fraction. Note the sizeable jump in the layer thickness at the smectic A-to-smectic B phase transition. Furthermore, the layer thickness decreases as a function of the volume fraction in for the smectic B and crystal phases. For the smectic A phase, there is a change in the layer thickness behaviour according to the aspect ratio and flexibility.}%
 \end{figure}

Perhaps the most interesting structural feature is how the smectic layer thickness depends on the contour length and persistence length of the particles. This is shown in Figure \ref{LayerN17}. For all cases, we find that transitioning from the smectic A to the smectic B phase, the layer spacing \emph{increases}. We speculate that this is due to the larger degree of in-layer packing possible in the more strongly ordered smectic B phase. In essence, this is caused by an increase in free volume. Depending on aspect ratio and flexibility, we observe that the layer spacing in the smectic A phase itself may in- or decrease with increasing concentration. This is not so for the smectic B and crystalline phases. It seems that in the smectic A phase, increasing the particle density may translate into a more or less proportional increase in in-layer density. If the in-layer density increases more strongly than the average density, then the layer spacing must increase. Because of the appreciable scatter in the data, we have not been able to find a clear trend. We also have no explanation for this phenomenon. 
 
 We notice that measurements of the concentration dependence of the layer thickness of smectic A and smectic B phases of the fd virus show approximately linear decrease of the layer spacing with increasing concentration \cite{Grelet2014}. This indeed is what we obtain for our longest particles. The jumps in layer spacing that we find at the smectic A-to-smectic B phase transition are not observed in the experiments on fd viruses. In addition, the actual layer spacings scaled to particle lengths are also smaller for fd viruses than for the particles in our simulations. On the other hand, we should not expect quantitative agreement with measurements on fd virus solutions on account of their much larger aspect ratio. 
 
 \section{Conclusion}
 \label{sec:con}

We preform molecular dynamics simulations to study the influence of flexibility and aspect ratio on the phase behaviour of purely repulsive, rod-like particles. Our particles have aspect ratios between 6 and 11, and ratios of the contour length over the persistence length between 0.05 and 0.5, \emph{i.e.}, we cover the range from very stiff to slightly flexible particles. By measuring the nematic, smectic and bond-order parameters and analysing correlation functions, we are able to distinguish five different phases. In order of increasing volume fraction these include isotropic, nematic, smectic A and B, and crystal phases. Of those phases we probe the structure of the particles and their arrangement, in particular in the smectic and crystalline phases.

In agreement with theoretical predictions and previous simulations, we conclude that the isotropic-nematic and nematic-smectic A phase transitions are sensitive functions of the aspect ratio and the flexibility of the particles. For the former, the larger the aspect ratio, the lower the volume fraction at the transition. For the former and the latter, the larger the flexibility, the larger the volume fraction at the transition. In fact, the smectic A phase disappears for sufficiently large of the ratio of the contour length over the persistence length, which is a measure for the bending flexibility of the particles. The transitions to the other, more highly ordered phases we find to be much less influenced by both the aspect ratio and the flexibility of the particles.

On increasing the concentration and going from the isotropic phase through the various liquid-crystalline phases to the crystal phase, we find that the end-to-end distance of the particles increasingly approaches their contour length. This is not entirely surprising because the more strongly ordered the phase, the more bending fluctuations are suppressed. In fact, we find, at least for our model bead-chain particles, that bending fluctuations are essentially completely suppressed in smectic B and crystalline phases, explaining the insensitivity of their stability to the persistence length. In other words, the particles in those phases are stretched to their contour length and resemble rigid rods.
  
The layer spacings that we find in the smectic A, smectic B, and crystal phases exceed the contour length of the particles. Interestingly and counter intuitively, these layer spacings need not decrease with increasing concentration of particles, at least in the smectic A phase. We find that depending on aspect ratio and flexibility spacing may actually increase. This is possible provided the increasing concentration is more than compensated by the in-layer density increase. We have not been able to pinpoint under what conditions this happens and also do not have an explanation for this phenomenon. Connected to this, we also find that the layer spacing increases upon going from the smectic A to the smectic B phase. This arguably is caused by the increase in free volume across the transition. The layer spacing in the smectic B and crystalline phases does behave as expected, that is decrease with increasing concentration.  

If we compare our phase diagrams with that of fd viruses in aqueous solution then all phases are reproduced, except for the columnar phase that for fd viruses occurs for concentrations in between the smectic B and crystal phases are stable. Of course, our particles are much shorter and perhaps it is that that suppresses the columnar phase in our simulations. The existence of the columnar phase in dispersions of monodisperse rod-like particles remains somewhat enigmatic and has been subject of a lot of debate in the literature \cite{Wensink2007}. It has been suggested that explicit modelling of the electrostatics stabilises that phase albeit that we cannot exclude the possibility that it is a question of a combination of flexibility and large-enough aspect ratio \cite{vdSchoot1996}. The challenge is to reach aspect ratios large enough to investigate this hypothesis.  


%

\begin{acknowledgments}

This project has received funding from the European Union's Horizon 2020 research and innovation programme under the Marie Sklodowska-Curie Grant Agreement No 641839 and from HFSP under Grant Agreement No RGP0017/2012.

\end{acknowledgments}

\begin{appendix}
\section{Onsager theory of linearly compressible hard rods}
\label{appendx}

We observe in our simulations the contraction of the average contour length of the chains for increasing volume fractions on account of their finite extensional compressibility. At the isotropic-nematic phase transition there is also a discontinuity in their contour length, with the particles in the nematic phase being slightly longer. A similar discontinuity occurs at the phase transition towards the smectic-B-crystal phase. These two observations can be rationalised with theoretical predictions we obtain by applying Onsager theory to extensible rods.

To this end we consider a system of $N$ bead-spring chains in a volume $V$ at temperature $T$. Each rod consists of $n$ beads connected with $n-1$ harmonic bonds with elastic constant $\kappa$ and rest length $r$. Each chain has total rest length $L_0=(n-1)r$ and diameter $D$. The contour length $L$ changes with the concentration of the dimensionless concentration $c=B_2^{\mathrm{iso}}\rho$ with $\rho=N/V$ the number density of particles. The free energy $F$ can be written as a function of the orientational distribution function $f(\Omega)$ and the compression of the chain $x=L/L_0$:
\begin{equation}
\frac{F[f]}{Nk_BT}=A+\ln{c}+\sigma[f]+c\rho[f]x^2+K(x-1)^2,
\end{equation}
where A is a constant, $\ln{c}$ is the ideal gas distribution, $\sigma[f]$ is the orientation entropy, $c\rho[f]x^2$ is the packing entropy, and $K(x-1)^2$ is related with the potential energy of the harmonic springs with
\begin{equation}
K=\frac{\kappa{L_0^2}}{2(n-1)k_BT}.
\end{equation}
The third and fourth terms mentioned previously are given by the expressions:
\begin{equation}
\sigma[f]=\int{f(\Omega)\ln{(4\pi f(\Omega))}d\Omega}
\end{equation}
and
\begin{equation}
\rho[f]=\frac{4}{\pi}\int{|\sin{\gamma}|f(\Omega){f(\Omega^\prime)}d\Omega{d}\Omega^\prime},
\end{equation}   
where $|\sin{\gamma}|$ is the angle between the chains with orientation $\Omega$ and $\Omega^\prime$.

In the isotropic phase, the normalised distribution function is $f(\Omega)={1}/{4\pi}$, resulting in an orientational entropy $\sigma[f]=0$ and a packing entropy $\rho[f]=1$. The free energy for the isotropic state becomes
\begin{equation}
\frac{F^{\mathrm{iso}}}{Nk_BT}=A+\ln{c}+cx^2+K(x-1)^2.
\end{equation}   
The equilibrium condition for $x$, $\frac{\partial}{\partial{x}}\left[\frac{F^{\mathrm{iso}}}{Nk_BT}\right]=0$, leads to the compression in the isotropic phase
\begin{equation}\label{xISO}
x^{\mathrm{iso}}=\frac{K}{c+K}.
\end{equation}  

For the nematic phase, we follow a similar procedure as Odijk \cite{Odijk1986}. We assume the orientational distribution function to be Gaussian and obey cylindrical and inversion symmetry: 
\begin{equation}
f(\theta) = \begin{cases} \alpha/4\pi\,\exp\left({-\alpha\theta^2/2}\right), & \mbox{if } 0{\leqslant}\theta\leqslant\pi/2 \\ \alpha/4\pi\,\exp\left({-\alpha(\pi-\theta)^2/2}\right), & \mbox{if } \pi/2<\theta\leqslant\pi \end{cases}, 
\end{equation}   
where the normalisation is only accurate for $\alpha\gg{1}$. For this distribution we have for the orientational entropy $\sigma[f]\sim\ln{\alpha}-1$ and for packing entropy $\rho[f]\sim4/\sqrt{\alpha\pi}$ \cite{Vroege1992}.
The free energy for the nematic state is then
\begin{equation}
\frac{F^{\mathrm{nem}}}{Nk_BT}=A+\ln{c}+\ln{\alpha}-1+\frac{4cx^2}{\sqrt{\alpha\pi}}+K(x-1)^2.
\end{equation}   
From this expression we find equilibrium values $\alpha=4c^2x^4/\pi$ and
\begin{equation}
\label{xNEM}
x^{\mathrm{nem}}=\frac{1}{2}+\sqrt{\frac{1}{4}-\frac{2}{K}}.
\end{equation}   
For $x^{\mathrm{nem}}$ there is also a negative root solution that we ignore for being physically unrealistic. For $K\rightarrow\infty$, $x^{\mathrm{nem}}=1$. For $K<8$, the compression becomes imaginary, meaning that the nematic phase becomes unstable.

Equations \ref{xISO} and \ref{xNEM} describe the behaviour of the mean length of our chains in the isotropic and nematic phases. We now calculate the coexistence concentration. Coexistence between two phases occur when the osmotic pressure $\Pi=-({\partial F}/{\partial V})_{N,T}$ and chemical potential $\mu=({\partial F}/{\partial N})_{V,T}$ are equal for both states, $\mu^{\mathrm{iso}}=\mu^{\mathrm{nem}}$ and $\Pi^{\mathrm{iso}}=\Pi^{\mathrm{nem}}$. From these equations we then calculate the coexistence concentrations for the isotropic $c^{\mathrm{iso}}$ and the nematic phase $c^{\mathrm{nem}}$.

We now add flexibility to our previous model to study how it affects the discontinuity in the average length of the chains at the isotropic-nematic phase transition. Our starting point is the expression derived by Odijk \cite{Odijk1985,Odijk1986} describing the orientational entropy for semi-flexible particles, $L/L_P\ll1$. For the isotropic phase there is no change of the orientational entropy. For the nematic phase there is the extra term $\sigma^{\mathrm{Odijk}}=L_0\alpha{x}/4L_P$, then the orientational entropy is 
\begin{equation}
\sigma=\ln\alpha-1+\frac{L_0\alpha{x}}{4L_P}.
\end{equation}

With this new orientational entropy, the free energy for the nematic phase becomes
\begin{equation}
\frac{F^{\mathrm{nem}}}{Nk_BT}=A+\ln{c}+\ln{\alpha}-1+\frac{L_0\alpha{x}}{4L_P}+\frac{4cx^2}{\sqrt{\alpha\pi}}+K(x-1)^2.
\end{equation}   
Solving the equilibrium value for $\alpha$ and $x$ we obtain the compression of the chain as a function of the dimensionless concentration. These and the coexistence concentrations are calculated numerically.

\begin{figure}
 \includegraphics[scale=.25]{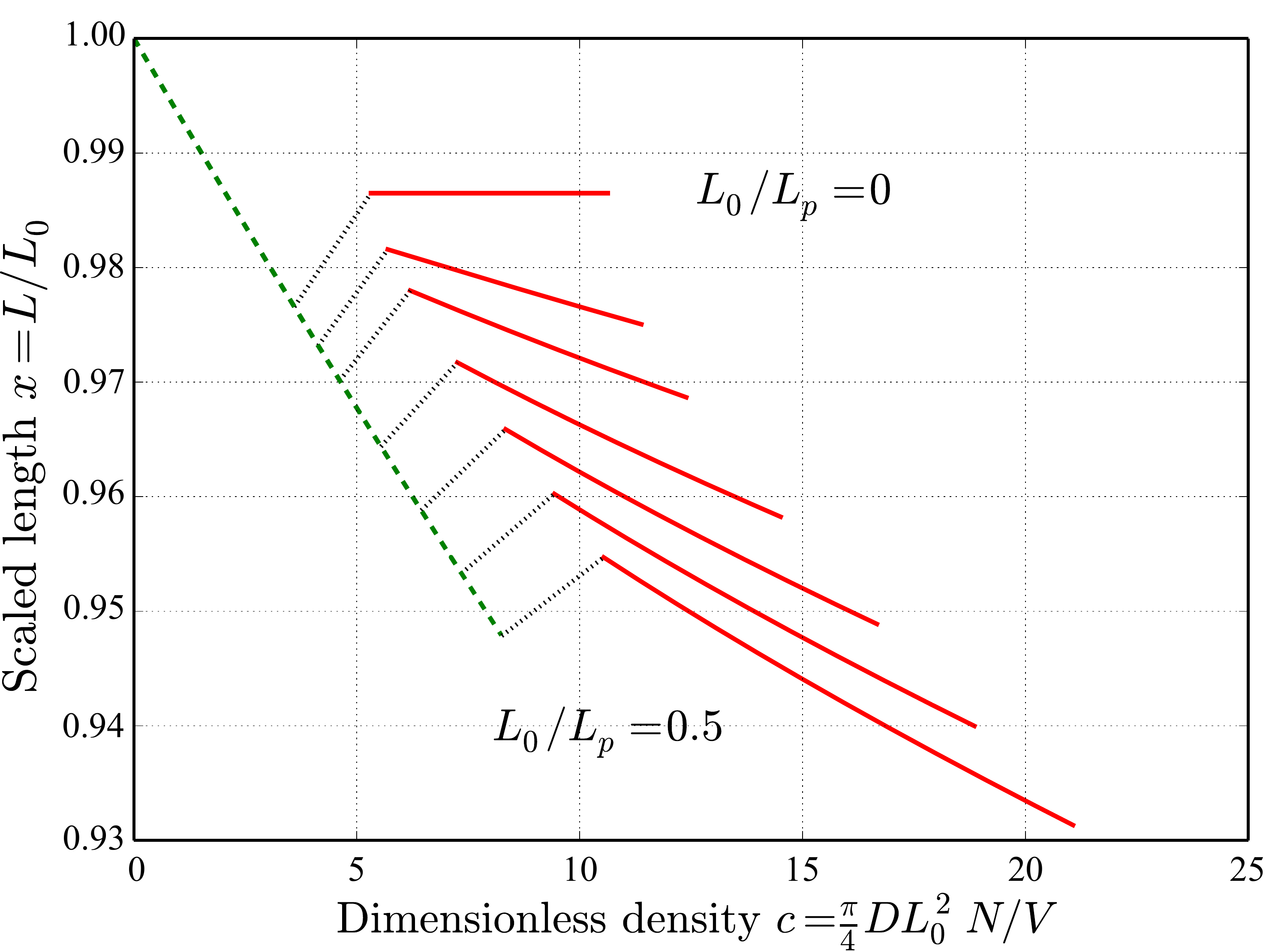}%
 \caption{\label{xc}Average length of the chains $x$ as a function of the dimensionless concentration $c$, for $K=150$ and flexibilities $L_0/L_P=0$, 0.05, 0.1, 0.2, 0.3, 0.4 and 0.5. The green dashed line represents the isotropic phase, the red lines represent the nematic phase, and the black dots connect the points of coexistence.}%
 \end{figure}

Finally, we compare the simulations with the model calculations. We specifically perform the calculations for $K=150$ and for flexibilities $L_0/L_P= 0$, 0.05, 0.1, 0.2, 0.3, 0.4 and 0.5, as can be seen in Figure \ref{xc}. These values coincide with the simulated values, except $L_0/L_P=0$, the rigid rod limit that we did not simulate. We find similarities between these results and our simulation results. First, the decrease of the average length of the chain with increasing concentration for both isotropic and nematic phase. Second, the discontinuity in the average length decrease with the increase of flexibility. There are some differences though, like the curvature of the lines, straight for the calculations and curved for the simulations. 

\end{appendix}


%

\end{document}